\documentclass[conference]{IEEEtran}
\usepackage{pifont}
\usepackage{xspace}
\usepackage[dvipsnames]{xcolor}
\usepackage{fontawesome}
\usepackage{enumitem}
\usepackage{multirow}
\usepackage{threeparttable}
\usepackage{natbib}
\usepackage{amssymb}
\usepackage[most]{tcolorbox} 
\tcbset{enhanced}            
\newtcbox{\code}{on line,
  colback=gray!15, colframe=gray!50,
  boxrule=0pt, arc=2pt, left=2pt, right=2pt, top=0.1pt, bottom=0.1pt,
  tcbox raise base, 
}

\setcitestyle{numbers,square,comma}

\usepackage{graphicx}
\usepackage{booktabs}
\PassOptionsToPackage{hyphens}{url}\usepackage{hyperref}

\newcommand{\mypar}[1]{\noindent \textbf{#1}.\xspace}

\newcommand{\namespace}{ns\xspace}
\newcommand{\Namespace}{NS\xspace}

\usepackage[available,functional,reproduced]{ndssbadges}
\begin{document}
\title{Breaking the Bulkhead: Demystifying Cross-Namespace Reference Vulnerabilities in Kubernetes Operators}

\author{\IEEEauthorblockN{Andong Chen\IEEEauthorrefmark{1},
Ziyi Guo\IEEEauthorrefmark{2}\ding{161},
Zhaoxuan Jin\IEEEauthorrefmark{2}, 
Zhenyuan Li\IEEEauthorrefmark{1}\ding{161} and
Yan Chen\IEEEauthorrefmark{2}}
\IEEEauthorblockA{\IEEEauthorrefmark{1}Zhejiang University \IEEEauthorrefmark{2}Northwestern University \\ chenandong@zju.edu.cn, \{n7l8m4, zhaoxuanjin2025\}@u.northwestern.edu, \\ lizhenyuan@zju.edu,cn, ychen@northwestern.edu} 
}

\IEEEoverridecommandlockouts
\makeatletter\def\@IEEEpubidpullup{6.5\baselineskip}\makeatother
\IEEEpubid{\parbox{\columnwidth}{
        \ding{161} Corresponding Authors: Ziyi Guo, Zhenyuan Li \\ \\
		Network and Distributed System Security (NDSS) Symposium 2026\\
		23 - 27 February 2026, San Diego, CA, USA\\
		ISBN 979-8-9919276-8-0\\  
		https://dx.doi.org/10.14722/ndss.2026.240761\\
		www.ndss-symposium.org
}
\hspace{\columnsep}\makebox[\columnwidth]{}}

\maketitle

\begin{abstract}
Kubernetes Operators, automated tools designed to manage application lifecycles within Kubernetes clusters, extend the functionalities of Kubernetes, and reduce the operational burden on human engineers.
While Operators significantly simplify DevOps workflows, they introduce new security risks. 
In particular, Kubernetes enforces namespace isolation to separate workloads and limit user access, ensuring that users can only interact with resources within their authorized namespaces. However, Kubernetes Operators often demand elevated privileges and may interact with resources across multiple namespaces. This introduces a new class of vulnerabilities, the Cross-Namespace Reference Vulnerability. 
The root cause lies in the mismatch between the declared scope of resources and the implemented scope of the Operator’s logic, resulting in Kubernetes being unable to properly isolate the namespace.
Leveraging such vulnerability, an adversary with limited access to a single authorized namespace may exploit the Operator to perform operations affecting other unauthorized namespaces, causing Privilege Escalation and further impacts.

To the best of our knowledge, this paper is the first to systematically investigate Kubernetes Operator attacks. 
We present Cross-Namespace Reference Vulnerability with two strategies, demonstrating how an attacker can bypass namespace isolation. 
Through large-scale measurements, we found that over 14\% of Operators in the wild are potentially vulnerable. 
Our findings have been reported to the relevant developers, resulting in 8 confirmations and 7 CVEs by the time of submission, affecting vendors including Red Hat and NVIDIA, highlighting the critical need for enhanced security practices in Kubernetes Operators. To mitigate it, we open-source the static analysis suite and propose concrete mitigation to benefit the ecosystem.
\end{abstract}

\IEEEpeerreviewmaketitle

\section{Introduction}
Kubernetes has emerged as the dominant platform for container orchestration, playing a central role in the deployment, scaling, and management of containerized applications in modern cloud-native environments \cite{cncfreport, redhatreport,edgedelta,datahub}. As a highly extensible and open-source system, Kubernetes facilitates the automation of complex operations such as container deployment, scheduling, and management across clusters. Its flexibility and wide adoption have made it the cornerstone of many enterprise-level infrastructure solutions, offering efficient ways to handle diverse and dynamic workloads in a scalable manner.

Kubernetes organizes resources into namespaces \cite{namespace}, which allow users to divide a single cluster into multiple virtual clusters. Each namespace serves as a logical boundary, isolating resources like pods, services, and secrets from other namespaces within the same cluster. This isolation is essential for managing different applications or services within the same Kubernetes environment, enabling users to work independently without interfering with each other. Namespaces also provide a way to scope access to resources, ensuring that certain actions can be confined to specific namespaces and reducing the risk of accidental or malicious interference between services.

To achieve namespace isolation, a crucial security mechanism is Role-Based Access Control (RBAC) \cite{rbac}. RBAC defines roles and permissions for users, service accounts, and other entities within the cluster, helping to control which actions are allowed within the system. For example, a staff member of a team may only be allowed to manipulate resources within the namespace assigned to their team, while the cluster administrator would be assigned all permissions across namespaces. By configuring RBAC policies, administrators can limit access to resources within specific namespaces, ensuring that users or services can only interact with the resources they are authorized to access. This granularity of access control reinforces the isolation between namespaces and helps to prevent unauthorized access to sensitive resources.

While Kubernetes provides robust tools and security mechanisms to manage and secure applications, the native platform has limitations in automating the lifecycle of complex applications. Kubernetes requires significant manual intervention for tasks like scaling, upgrades, and configuration management, which can be time-consuming and error-prone \cite{vmwarereport}. Kubernetes Operators \cite{operator} were introduced to address these limitations. Operators are programs that extend Kubernetes' capabilities by automating the management of applications. They encapsulate the operational knowledge required to manage complex Kubernetes applications, automating critical tasks such as deployment, scaling, and lifecycle management. Users can then easily request Operators to conduct complex operational tasks in contrast to manually manipulating raw Kubernetes resources. By automating these processes, Operators reduce the operational burden on DevOps teams and enable more consistent and reliable application management.

However, Kubernetes Operators require significant privileges to carry out their tasks. Due to the broad range of operations they need to perform, these Operators are usually granted substantial permissions across namespaces. While these permissions are necessary for the proper functioning of the Operator, they also introduce a significant security risk. Due to improper security practices in Operator implementation, adversaries may forge malicious requests toward Operators, exploit vulnerabilities to escalate their own permissions, break namespace isolation, and perform unauthorized operations within the cluster. 

Although extensive research has been conducted on Kubernetes security, Operator-specific attacks remain largely unexplored. Previous studies have focused on misconfiguration in Kubernetes, especially excessive RBAC permissions, highlighting the risks of overly permissive access controls \cite{takeover, epscan, podescape, k8sblackhat}. Practitioners thus suggested limiting RBAC permissions for Kubernetes Operators \cite{operatorsecpractice, operatorsynk, operatorkubeops}. While these works have led to improvements in reducing permissions, they have neither addressed the inherent risks that remain even when permissions are minimized, nor presented attacks specific to Operators. Specifically, necessary permissions required by Operator business logics, which cannot be further minimized, may still be exploited due to improper security design within the Operator logic. Furthermore, existing attacks assume attackers have compromised containers (e.g., get a shell) in prior, leaving critical gaps in how to compromise application containers.

Other research has focused on bugs in Kubernetes Operators \cite{operatorbug, sieve, acto}, yet these studies primarily concentrated on the functional bugs of Operators rather than their security vulnerabilities. Furthermore, existing security tools \cite{trivy, kubescape, kubesec, kubearmor, opa, kyverno} do not adequately address the security concerns specific to Operators, leaving a significant gap in the ecosystem.

Thus, in this paper, we present the first systematic research on Kubernetes Operator attacks, unveiling cross-namespace reference vulnerabilities. 
The root cause of cross-namespace reference vulnerability lies in the mismatch between the declared scope of a resource and the effective scope of the Operator's logic. A resource may be declared as namespace-scoped, allowing users with limited access to a single namespace to deploy it, while the Operator’s logic may perform actions that affect other namespaces, breaking the intended isolation. We present a novel practical threat model without assuming that attackers have already compromised containers, allowing exploits to be made from scratch.
We propose two distinct tactics for exploiting Kubernetes Operators to elevate an attacker's privileges, both of which exploit the scope mismatch in Operator implementation and break the isolation between Kubernetes namespaces. 


To measure the new kind of vulnerabilities, we designed and implemented a static analysis suite that can identify scope mismatch in Operators. We conducted large-scale measurements of 2,268 Kubernetes Operators in the wild, revealing that over 14\% of the Operators are potentially vulnerable to these attacks. We responsibly disclosed our findings to their developers, and, by the time of submission, 8 vulnerabilities had been confirmed and 7 CVEs were assigned or under assignment in response to our reports, affecting vendors Red Hat and NVIDIA, highlighting the critical need for enhanced security practices in Kubernetes Operators.

All in all, our contributions can be summarized as follows:

\begin{itemize}[leftmargin=\parindent]
\item \mypar{New Attack} We present the first systematic research on Kubernetes Operator attacks. 
We unveil the new type of vulnerability specific to Operators, Cross-Namespace Reference Vulnerability, detailing two distinct tactics and their root cause, both of which can be leveraged to escalate privileges.
\item \mypar{Large-scale Measurement} We design and implement tools for the measurement of real-world Cross-Namespace Reference Vulnerabilities. We conduct large-scale measurements of Kubernetes Operators in the wild, demonstrating that over 14\% of Operators are susceptible to these vulnerabilities. 
\item \mypar{Real-World Impact} We responsibly disclosed our findings to the developers, and, by the time of submission, 8 vulnerabilities were confirmed and 7 CVEs were assigned in response to our reports, affecting leading vendors including Red Hat and NVIDIA.
\item \mypar{Benefic Ecosystem} We will open-source our analyzer, which covers detailed modeling of major Kubernetes libraries, enabling analysis of not only Operators but also other Kubernetes applications. To facilitate remediation, we propose concrete mitigations, open-source mitigation samples, and a patch generator for practitioners. 
\end{itemize}
\section{Background}

\subsection{Kubernetes Namespace and RBAC}

Kubernetes is a powerful container orchestration platform that automates the deployment, scaling, and management of containerized applications. It is designed to manage large-scale complex applications, where multiple teams or applications may share a single cluster. To help organize and isolate resources within the cluster, Kubernetes provides a mechanism called \textit{Namespaces} \cite{namespace}. A namespace\footnote{We use \namespace as short for namespace for the remaining parts.} is a logical partition or a virtual cluster within a physical cluster. Each namespace acts as a boundary, ensuring that resources in one namespace do not conflict with those in another.

Namespaces are particularly useful in multi-tenant environments, where different teams or applications share the same Kubernetes cluster \cite{kubernetes-multitenancy}. By isolating resources in separate namespaces, Kubernetes prevents one team from accessing or interfering with another team's resources. This isolation is vital for security and resource management, ensuring that users and applications can only access the resources assigned to their namespace, preventing unauthorized access or potential conflicts between resources.

Kubernetes employs multiple security mechanisms to help ensure that the cluster remains secure and that resources are properly isolated. One of the most important mechanisms for securing access to resources within a namespace is RBAC (Role-Based Access Control) \cite{rbac}. RBAC allows administrators to define roles with specific permissions and bind those roles to users or service accounts, ensuring that only authorized entities can perform certain actions. Cluster administrators can grant both \namespace-specific permissions and cluster-level permissions. Specifically:
\begin{itemize}[leftmargin=\parindent]
\item \textit{Role} and \textit{RoleBinding}: Define and grant resource permissions within a specific namespace to a user, group, or service account.
\item \textit{ClusterRole} and \textit{ClusterRoleBinding}: Define and grant resource permissions across all namespaces to a user, group, or service account.
\end{itemize}

\subsection{Kubernetes Resource}
At its core, Kubernetes organizes the cluster's state using resources, which are data objects encapsulating configuration and runtime information.
Kubernetes manages many types of resources within a cluster, which are fundamental components that define the desired state of applications and services. Common built-in resource types include pods \cite{pod} and deployments \cite{deployment}. A pod is the smallest deployable unit in Kubernetes and typically represents one or more containers that share the same network and storage resources. A deployment is a higher-level abstraction that manages the lifecycle of pods, specifying the desired number of pod replicas.
In addition to built-in resources, Kubernetes allows users to define \textit{Custom Resources} (CRs) \cite{cr}, extending Kubernetes to manage domain-specific requirements beyond its default capabilities. Each type of Custom Resource is described by a Custom Resource Definition (CRD) \cite{cr}, which specifies the resource's schema.

In Kubernetes, each type of resource, whether a built-in resource or a Custom Resource, is bound with an explicit scope, indicating its accessibility and impact within the cluster. Resources can be either \textit{Namespace-scoped} or \textit{Cluster-scoped}. In Kubernetes, the scope of built-in resources is embedded within the Kubernetes implementation, whereas the scope of Custom Resources is explicitly defined in their associated Custom Resource Definitions. \Namespace-scoped resources must reside within a specific namespace, meaning they are logically isolated and can be accessed or manipulated by users with only \Namespace-specific Roles. 
Conversely, Cluster-scoped resources exist at the cluster level and are not confined to any single namespace. These Cluster-scoped resources may affect or interact with all namespaces across the cluster. Due to their broad impact, accessing or manipulating cluster-scoped resources requires users to possess a cluster-wide ClusterRole, reflecting elevated privileges.


Importantly, resources themselves merely represent desired configurations or states. To realize these desired states, each type of resource is managed by an associated controller, a program responsible for monitoring resources and taking actions to align the actual state with the desired state described by resources. For example, the \textit{deployment controller} monitors \textit{deployment} resources and ensures that the desired number of pod replicas are running. If a pod fails or is deleted, the \textit{deployment controller} automatically creates a new pod to meet the desired state.
For built-in resources, Kubernetes provides native controllers. For Custom Resources, users should develop custom controllers.

\subsection{Kubernetes Operator}
Kubernetes Operator is a method of automating and managing the lifecycle of complex applications on top of Kubernetes by extending the platform’s native capabilities. Originally introduced by CoreOS (now part of Red Hat), it emerged from a recognition that while Kubernetes excels at automatically orchestrating workloads, many organizations need a more powerful automation pattern to handle full lifecycle management, such as database management, application upgrades, or failure recovery, that requires specific operational knowledge. 
Operator is thus introduced to extend Kubernetes by embedding human operational expertise into software, enabling automated management of complex, stateful applications.

An Operator consists of one or more Custom Resource Definitions (CRDs) and their corresponding Custom Resource Controllers. CRD defines the schema of a custom resource type that will be processed by the Operator. Controllers work with CRDs by continually monitoring custom resources and taking actions to fulfill operational tasks requested by users.









To use an Operator, users manipulate custom resources that represent the operational task, along with the related parameters they want to conduct. The Operator controller reads these custom resources, takes actions listed in the custom resource, and ensures that the operational task behaves as expected. 
\begin{figure}
    \centering
    \includegraphics[width=\linewidth]{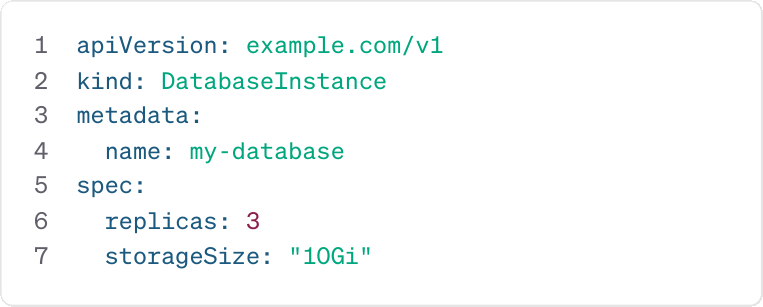}
    \caption{Custom Resource Example}
    \label{fig:cr-example}
\end{figure}
Considering an Operator for database management tasks, users may create a custom resource listed in \autoref{fig:cr-example}, including arguments like the number of database replicas and storage settings. The Operator controller then reads the custom resource, automatically provisions, scales, and maintains the database according to these specifications.



Since Operators typically manage multiple kinds of resources across namespaces, they often run with elevated RBAC privileges, allowing them to create, modify, and delete resources on behalf of users. This makes them powerful but also introduces security risks. If an Operator does not adopt proper security measures, attackers may exploit the vulnerabilities of the Operator to gain unauthorized access or manipulate resources beyond their intended scope.

\section{Threat Model}
\label{sec:threatmodel}

Our threat model aligns with real-world Kubernetes deployments where multiple tenants, teams, or applications share the same cluster while being isolated within their respective namespaces \cite{kubernetes-multitenancy}.
The adversary aims to break Kubernetes namespace isolation and achieve cross-\namespace privilege escalation by exploiting security weaknesses in Operator implementations. Their objectives are performing operations in unauthorized namespaces (i.e., namespaces that they have no \textit{Roles}) and thus escalating privileges.

We assume the Kubernetes cluster deploys vulnerable Operators, and the adversary has legitimate access to a Kubernetes cluster but can only access their authorized namespaces. Thus, they cannot access or manipulate cluster-scoped resources and can only interact with Operators by manipulating \namespace-scoped resources in their authorized namespace. They seek to leverage vulnerable Kubernetes Operators to execute unauthorized operations in other namespaces.
The adversary may be:
\begin{itemize}

\item A malicious tenant in a multi-tenant cluster who is only authorized to access their assigned namespace.

\item A compromised application running in a namespace with \namespace-level permissions mounted.

\item An attacker who steals credentials of Kubernetes accounts with \namespace-level permissions.
\end{itemize}

Our threat model is practical. For example, many real-world cloud services are operated on multi-tenant Kubernetes. One of them, Red Hat Developer Sandbox \cite{developersandbox}, is provided by assigning a namespace on multi-tenant Kubernetes to a user, where Operators are deployed. This implies that attackers may subscribe Kubernetes-based service to conduct our attacks. 

It is notable that our threat model is significantly different from previous works~\cite{takeover, epscan, podescape, k8sblackhat}. Specifically, existing works assume that the adversaries have compromised vulnerable application containers in prior, which is a strong assumption in the real world, leaving critical gaps in how to compromise application containers. In contrast, within our threat model, adversaries don't have to gain control of Operators in prior, since these Operators may be deployed in adversary-unauthorized namespaces. In extreme cases, Operators can even be deployed outside the Kubernetes cluster \cite{operatoroutside}. So the threat model of previous works is relatively infeasible, but our threat model is more feasible and aligned with real-world scenarios.

In this paper, the terms \textit{Namespace} and \textit{Cross-Namespace} refer specifically to Kubernetes Namespaces, which are used to isolate resources within a Kubernetes cluster. They are distinct from similarly named concepts in other systems, such as Linux Namespaces, which isolate system resources at the OS kernel level. They isolate resources at different layers and have no direct linkage despite sharing the same term.

The vulnerability we presented acts as a strategy for attackers to access or manipulate unauthorized resources. The final concrete impact depends on the accessed unauthorized resources and differs between cases. 

\section{Cross-Namespace Attacks}
\label{sec:attacks}

\begin{figure*}[t]
    \centering
    \includegraphics[width=0.93\linewidth]{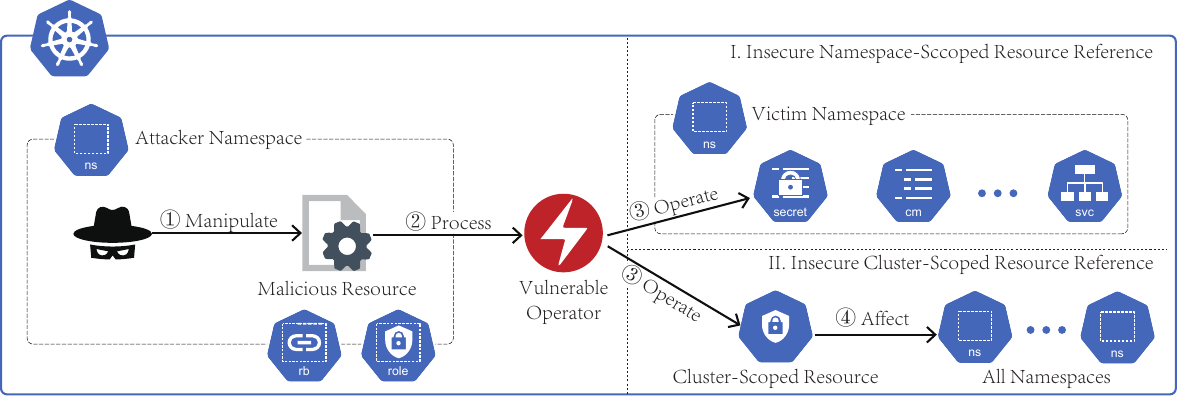}
    \caption{Attack Flow}
    \label{fig:attack-flow}
\end{figure*}

\subsection{Attack Overview}
In Kubernetes clusters, namespaces act as virtual boundaries, restricting user access and isolating resources. Kubernetes Operators manage applications and resources and perform essential operational tasks. While these Operators simplify application management, their inherent privileges and operational flexibility create potential security vulnerabilities that can be exploited for cross-namespace reference attacks.

The high-level attack flow is as follows: an attacker, who has legitimate but restricted access to one namespace, manipulates a maliciously crafted \namespace-scoped resource instance within their authorized namespace. The Operator, continuously watching for \namespace-scoped resource events, detects this malicious resource instance and processes it with privileged operations that impact namespaces beyond the attacker’s authorized scope, effectively breaking the intended namespace isolation enforced by Kubernetes.

\mypar{Root Cause}
The core enabling cross-\namespace reference attacks stems from a mismatch between the declared scope of a resource and the actual scope of its process logic. Specifically, the vulnerability arises when the scope of a resource is defined as \textit{Namespaced}, indicating that each instance should strictly reside within its assigned namespace. Thus, an adversary only with \textit{Role} in a single namespace is allowed by RBAC to manipulate such a resource in their own namespace. However, despite this \namespace-scoped definition, the Operator may actually perform operations across namespaces, inadvertently allowing manipulation of resources in namespaces beyond the intended scope. As a result, an adversary without \textit{Role} in other namespaces may invoke the Operator to escalate their permissions and access unauthorized namespaces.

The root leading to such implementation lies in the improper trade-off at the design level between security and convenience.
Kubernetes grants significant flexibility for controllers. Built-in controllers in Kubernetes prioritize security to eliminate attack surfaces. For instance, a Pod can only reference Secrets within the same namespace to avoid cross-\namespace. However, many developers \cite{kubeissue57325, kubeissue104415, kubeissue107495, wantcrossns, wantcrossns2} want cross-\namespace to avoid manually duplicating Secrets in each namespace. Since Operators are proposed to reduce human intervention, the community may prioritize convenience, overlooking the underlying Kubernetes’s security model and enabling cross-\namespace reference.

\mypar{Cross-Namespace Features} 
There are two primary scenarios that enable cross-namespace reference actions. First, when processing \namespace-scoped resources in one namespace, an Operator may access or manipulate other \namespace-scoped resources in a different namespace (\S \ref{subsection:cross-namespace-ref}). 
Second, when processing \namespace-scoped resources, an Operator might access or manipulate cluster-scoped resources, leading to impacts on the whole cluster across all namespaces (\S \ref{subsection:cluster-ref}).
Both scenarios allow adversaries to trick the Operator into performing unintended, privileged operations beyond the adversary's RBAC scope.

\subsection{Insecure Namespace-Scoped Resource Reference}
\label{subsection:cross-namespace-ref}

Insecure \Namespace-Scoped Resource Reference vulnerability arises when an Operator processing \namespace-scoped resources, the fields of which are then used by the Operator to reference resources in other namespaces. This vulnerability fundamentally undermines Kubernetes namespace isolation by enabling attackers to indirectly access resources from namespaces they are otherwise restricted from accessing.

\mypar{Attack Flow}
As illustrated in \autoref{fig:attack-flow}, consider two namespaces: an attacker namespace and a victim namespace containing sensitive resources. The Roles and RoleBindings claim that the attacker can only access resources in their namespace and cannot access those in the victim's namespace.

To bypass the restriction of RBAC and access resources in unauthorized namespaces, the attacker first crafts and deploys a malicious resource instance within their namespace. This resource includes fields leveraged by the Operator to reference resources located in the victim namespace. From the perspective of Kubernetes, the deployment of the malicious resource should be allowed because it only knows that the attacker has created a resource under their authorized namespace, but does not know if the resource leads to privilege escalation.

The Operator then processes the created malicious resource. The Operator extracts the fields in the malicious resource, operates the victim resource located in the victim namespace, and inadvertently leaks or tampers with sensitive information. Thus, the attacker effectively escalates their privileges, bypassing Kubernetes' namespace isolation, gaining unauthorized access to resources that should have remained secure.

\mypar{Example}
A common real-world scenario occurs when an Operator manages applications consuming credentials (e.g., API Secret Key) stored in Kubernetes Secrets. As per Kubernetes official security practice \cite{Goodprac92:online}, Secrets should only be referenced strictly within the same namespace to maintain proper isolation. 

\begin{figure}
    \centering
    \includegraphics[width=0.80\columnwidth]{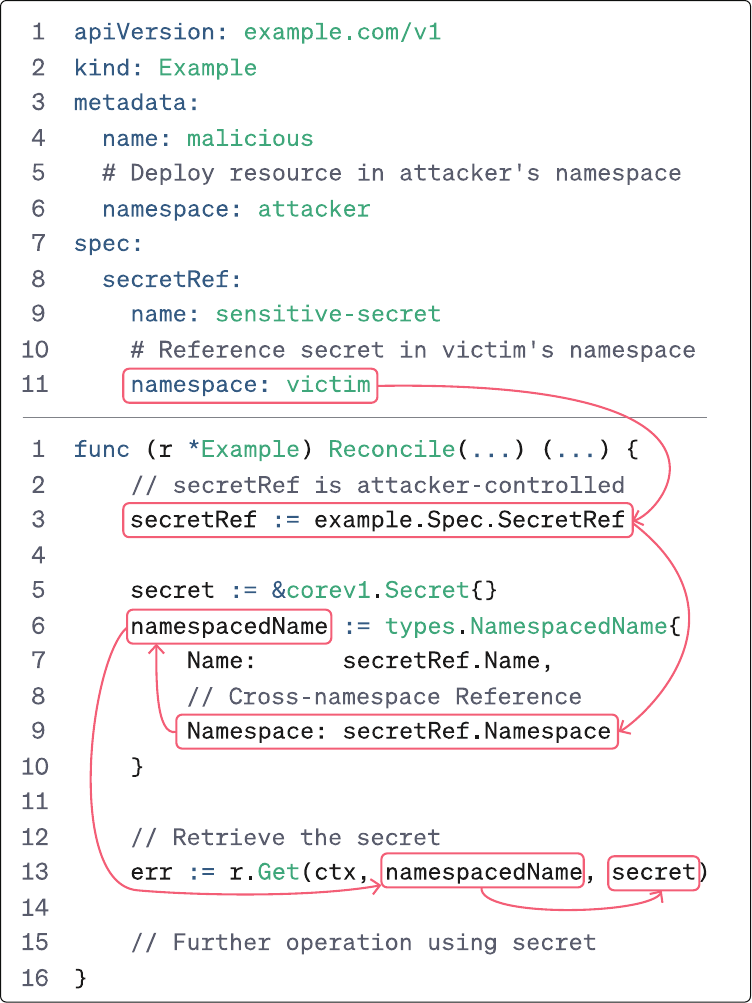}
    \caption{Insecure Namespace-Scoped Resource Reference Sample}
    \label{fig:cross-namespace-resource-sample}
\end{figure}

However, the vulnerable Operator implements cross-namespace references by setting up a \textit{secretRef.namespace} field in its custom resource definition.
Given this insecure implementation, an attacker restricted to a namespace could craft the malicious custom resource as illustrated in the upper YAML file of \autoref{fig:cross-namespace-resource-sample}. He deploys a resource with \textit{metadata.namespace} setting to \textit{attacker}, which means the resource is deployed in the \textit{attacker} namespace. This deployment is allowed since the RBAC authorized the attacker to work in his own namespace. In the specification of the resource, the attacker defines the value of \textit{secretRef} at Lines 8-11, referencing a Secret named \textit{sensitive-secret} in his unauthorized namespace \textit{victim}.

The Operator notices the malicious resource deployed by the attacker, reads the \textit{secretRef} field at Line 3 of the Reconcile function illustrated in \autoref{fig:cross-namespace-resource-sample}. The name and namespace of the referenced victim Secret are then loaded into the \textit{namespacedName} object, which is used to query and retrieve the specified Secret into the \textit{secret} object at Line 13. The remaining parts of the Operator will consume the content of the Secret to perform further operations.


\mypar{Impact}
Insecure \Namespace-Scoped Resource Reference vulnerabilities fundamentally enable attackers to escalate privileges by allowing them to reference and manipulate resources in namespaces beyond their legitimate access. 
Further impact of this vulnerability heavily depends on how the Operator processes and utilizes the referenced resources, as well as the nature of the referenced resources themselves. For instance, if the referenced resource is a Kubernetes Secret containing sensitive credentials like API Tokens, an attacker may obtain unauthorized access to applications, databases, or cloud infrastructure. If the Operator not only reads but also modifies referenced resources, attackers might disrupt service availability, modify application configurations, or inject malicious workloads. Thus, the severity and scope of the impact are highly context-dependent, ranging from sensitive information leakage to complete cluster compromise, based on the type and usage of the improperly referenced resource.

\subsection{Insecure Cluster-Scoped Resource Reference}
\label{subsection:cluster-ref}
Insecure Cluster-Scoped Resource Reference occurs when a Kubernetes Operator processes a \namespace-scoped resource and interacts with cluster-scoped resources. Unlike \namespace-scoped resources that remain isolated within specific namespaces, cluster-scoped resources affect the entire Kubernetes cluster. If an Operator allows users to influence these cluster-scoped resources through \namespace-scoped resources, it creates a pathway for attackers to escalate privileges and potentially compromise the entire cluster.


\mypar{Attack Flow}
As illustrated in \autoref{fig:attack-flow}, consider a namespace controlled by an attacker named \textit{attacker}, and all the other victim namespaces. The Roles and RoleBindings in the cluster define that the attacker can only access resources in their namespace and cannot access any victim's namespace.

The basic attack workflow for this vulnerability starts with an attacker creating a malicious resource in their authorized namespace, whose fields are leveraged by the Operator to reference a cluster-scoped resource. 
Kubernetes RBAC accepts the deployment of the malicious resource because it only knows that the attacker has created a resource under their authorized namespace, but does not know if the resource leads to privilege escalation.
The Operator, running with elevated cluster-level privileges, processes this malicious input and subsequently performs operations on the referenced cluster-scoped resource. As cluster-scoped resources inherently affect the entire Kubernetes environment, these unauthorized accesses and manipulations enable attackers to escalate privileges beyond their initial namespace boundaries and gain control or influence over all the other namespaces.

\mypar{Example}
\begin{figure}
    \centering
    \includegraphics[width=0.80\columnwidth]{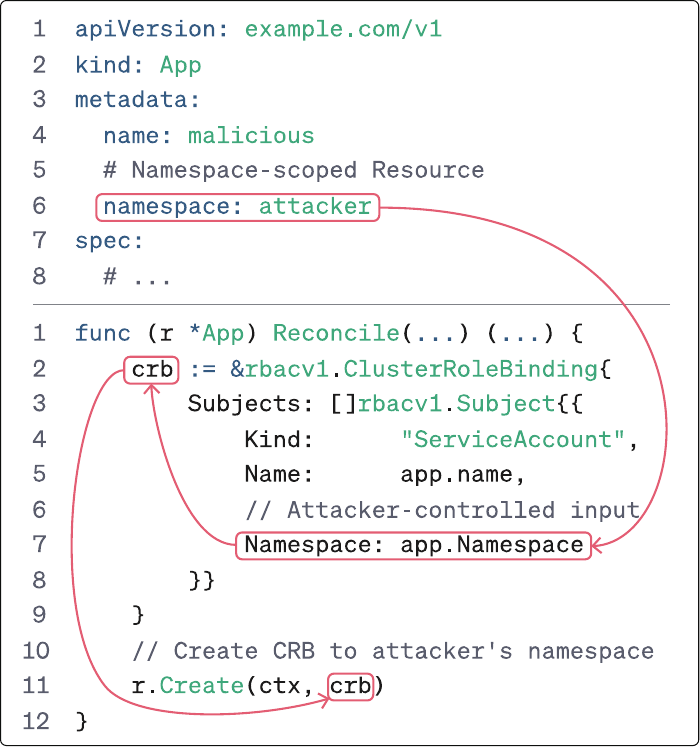}
    \caption{Insecure Cluster-Scoped Resource Reference Sample}
    \label{fig:cluster-scoped-sample}
\end{figure}
ClusterRole and ClusterRoleBinding are two cluster-scoped built-in resources in Kubernetes. It can grant cluster-wide permissions to a ServiceAccount, a built-in namespace-scoped resource that provides an identity for applications to access the Kubernetes API. 
Some Kubernetes applications require mounting a ServiceAccount with cluster-wide permissions to access the Kubernetes API and operate correctly. 
Thus, when deploying such an application in a specified namespace, their Operators will create a ServiceAccount in that namespace, mount it on the application, and create a ClusterRoleBinding to grant cluster-wide permission to that ServiceAccount.
An insecure implementation occurs when an Operator accepts \namespace-scoped resources and creates a ClusterRoleBinding to a ServiceAccount in the requesting namespace. 
An attacker restricted to a namespace could thus craft the malicious resource illustrated in the YAML file of \autoref{fig:cluster-scoped-sample}. He deploys a resource with \textit{metadata.namespace} setting to \textit{attacker}, which means the resource is deployed in the \textit{attacker} namespace.

The Operator monitors \textit{App} custom resources. It finds the malicious resource deployed by the attacker, creates a ServiceAccount in the attacker's namespace, and then creates a ClusterRoleBinding towards that ServiceAccount to grant it cluster-wide permissions. Since the ServiceAccount is created in the attacker's namespace, the attacker can impersonate the ServiceAccount to escalate his privilege, gaining cluster-wide permissions granted by ClusterRoleBinding. In short, such a vulnerability can be leveraged to directly elevate the permissions of attackers.

\mypar{Impact}
Insecure Cluster-Scoped Resource References allow attackers to escalate privileges and affect resources across all namespaces in a cluster. 
The specific severity and effect of this vulnerability depend on which cluster-scoped resources the Operator interacts with. For instance, if an Operator insecurely creates a ClusterRole or ClusterRoleBinding as dictated by \namespace-scoped resources, an attacker can gain cluster-level permissions. The detailed permissions assigned depend on the implementation of Operators.
\section{Cross-Namespace In The Wild}

To assess the prevalence of the vulnerabilities described in Section \ref{sec:attacks}, we conducted a large-scale measurement of real-world Kubernetes Operators and disclosed our findings to affected vendors.

\subsection{Measurement Methodology}

\subsubsection{Overview}

To understand how widespread the vulnerabilities are in real-world Kubernetes Operators, we perform a systematic measurement illustrated in \autoref{fig:measurement-flow}, which consists of the following steps:

\begin{figure*}[ht]
    \centering
    \includegraphics[width=\linewidth]{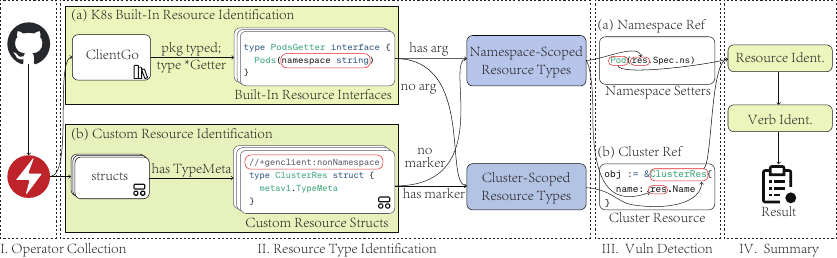}
    \caption{Measurement Workflow}
    \label{fig:measurement-flow}
\end{figure*}

\begin{enumerate}[leftmargin=15pt]
\item Operator Collection:
A large set of publicly available Kubernetes Operator repositories is collected from GitHub.


\item Resource Type Identification:
Resource types (either Kubernetes built-in resources or custom resources) used by each Operator are extracted, and their declared scopes (either \namespace-scoped or cluster-scoped) are identified based on the code.


\item Vulnerability Detection:
The analysis identifies whether Operators process \namespace-scoped resources but conduct insecure cross-\namespace reference behavior, as depicted in Section \ref{subsection:cross-namespace-ref} and Section \ref{subsection:cluster-ref}.

\item Summary:
Identified vulnerabilities are further aggregated based on the types of referenced resources and operation verbs to evaluate the impacts in the real world.
\end{enumerate}

Among all Kubernetes development frameworks recommended by the Kubernetes official \cite{operator}, Golang frameworks (i.e. Kubebuilder \cite{kubebuilder} and Operator SDK \cite{operatorsdk}) own the highest GitHub Stars and dominates with around 2.4k Operators, followed by Python (331 Operators), Shell (167 Operators), Java (134 Operators), Rust (96 Operators), and .NET (29 Operators). Given the overwhelming amount of Golang-based Operators, the collection specifically targets Operators implemented in Golang. We adopted CodeQL \cite{CodeQL} v2.17.4 to analyze Operators. The entire CodeQL query suite uses around 1,500 lines of QL rules, covering detailed modeling of major Kubernetes Go libraries, enabling analysis of Kubernetes applications beyond Operators. 

\begin{table}[ht]
\centering
\caption{Common Operator-Related Libraries}
\label{tab:modeled-library}
\begin{tabular}{@{}ll@{}}
\toprule
\textbf{Library}             & \textbf{Description}                                       \\ \midrule
k8s.io/api \cite{k8sapi}         & K8s Built-In Resource Specifications   \\
k8s.io/apimachinery \cite{apimachinery} & K8s Metadata Specifications \\
client-go \cite{clientgo}          & K8s Official Client           \\
client-gen \cite{gen}          & K8s Official Client Generator                       \\
controller-runtime \cite{controllerruntime} & Controller Client                             \\
\bottomrule
\end{tabular}
\end{table}

To enhance the measurement process, 5 commonly used libraries listed in \autoref{tab:modeled-library} were modeled to accurately resolve and track Kubernetes interactions within collected Operator implementations. They contain specifications of native Kubernetes resources, namespace-related data structures, and functions for Operators to manipulate resources. The detail is elaborated later.

\subsubsection{Operator Collection}
The dataset of Kubernetes Operators analyzed was collected by crawling GitHub repositories. To achieve this, GitHub Search API \cite{githubsearch} was utilized with the query string \textit{``Kubernetes Operator language: go"}. The collection process strictly adhered to GitHub's API usage policies to responsibly retrieve relevant Operator repositories.

After collecting Operators from GitHub, we set up CodeQL databases for each Operator. 
13 Operators that cannot be compiled to generate the CodeQL database due to errors, like syntax and dependency errors, are eliminated, and the final set contains 2,268 Operators.



\subsubsection{Resource Type Identification}
The first critical step in detecting vulnerabilities is Resource Type Identification, as the attack requires the attacker to initiate operations using a \namespace-scoped resource they are authorized to create in their namespace.
The analysis separately handles Kubernetes Built-in Resources and Operator-defined Custom Resources.

\mypar{Custom Resource Identification}
For Custom Resources defined by the Operators, their data structures can be explicitly extracted from the source code. In Kubernetes, each resource structure must contain a field of type TypeMeta (defined by the \textit{Apimachinery} library \cite{apimachinery}), which acts as the unique identifier of a resource type. Thus, the analyzer extracts all struct types in Operators and filters those with TypeMeta fields. This outputs all custom resource types in Operators.

To further identify the scope of each resource type (\namespace or cluster), common Kubernetes frameworks like Kubebuilder \cite{kubebuilder} and Operator SDK \cite{operatorsdk}, as well as Kubernetes’ official client code generator \cite{gen}, require developers to explicitly decorate cluster-scoped resource structs using special marker annotations ``\textit{+genclient:nonNamespaced}" or ``\textit{+kubebuilder:resource:scope:Cluster}".
By detecting these markers in the Custom Resource struct definitions, the analyzer reliably identifies Custom Resource types in operators and their scopes.

\mypar{Kubernetes Built-in Resources}
Unlike Custom Resources, the built-in Kubernetes resource specifications are imported from the external \textit{k8s.io/api} \cite{k8sapi} library to Operators, thus their source code and scope markers are not directly accessible for CodeQL. Therefore, the analyzer adopts an alternative method. Specifically, Operators would ultimately depend on the \textit{client-go} library \cite{clientgo}. Each type of built-in Kubernetes resource is uniquely associated with a typed client provided by the \textit{client-go} library \cite{clientgo}. Each typed client is constructed by methods in its corresponding \textit{Getter} interfaces under the \textit{k8s.io/client-go/kubernetes/typed} package. For example, considering the built-in resource type \textit{Pod}, there is a uniquely associated Pod client. The Pod client is constructed by the only method in the \textit{PodGetter} interface. 
By extracting the return types of methods in all \textit{Getter} interfaces, the analyzer identifies all built-in types and their clients. 

To determine the scope of built-in resources, the analysis leverages the only method in the \textit{Getter} interface of each typed client. Specifically, \namespace-scoped resource clients in client-go require a \textit{namespace} parameter in their constructor to specify the target namespace. In contrast, constructors for cluster-scoped resource clients do not require such a namespace argument. By counting and verifying constructor parameters, the analysis distinguishes \namespace-scoped from cluster-scoped built-in resources.

\subsubsection{Vulnerability Detection}
This step determines whether an attacker-controlled input can influence sensitive operations that cross namespace boundaries or impact the whole cluster. To achieve this, the analysis uses interprocedural taint tracking, tracing the propagation of data originating from the \namespace-scoped resource objects to insecure reference sites in the controller logic.

\mypar{Insecure Namespace-Scoped Resource Reference} 
The goal of this detection is to determine whether attacker-controlled values can be used to specify the namespace of another resource by the Operator. This is essential because if the attacker can influence which namespace a referenced resource belongs to, they can trick the Operator into accessing or modifying resources beyond their authorized scope.

\begin{table}[ht]
\centering
\caption{Resource Namespace Setters}
\label{tab:namespace-setter}
\begin{tabular}{@{}ll@{}}
\toprule
\textbf{Field}                                  & \textbf{Library}             \\ \midrule
ObjectMeta.Namespace                   & k8s.io/apimachinery \\
NamespacedName.Namespace               & k8s.io/apimachinery \\
ObjectMetaApplyConfiguration.Namespace & client-go           \\ \midrule
                                       &                     \\ \midrule
\textbf{Function}                               & \textbf{Library}             \\ \midrule
ApplyConfiguration.WithNamespace()     & client-go           \\
*.SetNamespace()                       & k8s.io/apimachinery \\
Constructor of Typed Client            & client-go          \\
Constructor of Typed Client            & client-gen           \\ \bottomrule
\end{tabular}
\end{table}

Thus, the analysis tracks data flow from \namespace-scoped resource objects to namespace setters (listed in Table \ref{tab:namespace-setter}) that are leveraged to specify the namespace of a resource. By systematic code reviews of common Operator libraries, we identify 3 struct fields that can store namespace values in 2 libraries and identify 4 types of functions in 3 libraries that can be used to set the namespace field of a resource object or set up a typed client towards a specific namespace.

It is worth noting that the 3 fields (listed in Table \ref{tab:namespace-setter}) of \namespace-scoped resource objects are excluded from the taint source, as these fields denote the namespace where this resource is deployed. Since the attackers are only authorized to access their namespace, the namespace fields of attacker-controlled resources are always the attacker-authorized namespace. If these fields sink in the referenced resources' namespace fields, it means the referenced resources are also in the attacker-authorized namespace. Thus, no cross-\namespace operation is conducted.

Taking the above key points into consideration, if the tainted data flows into any of these namespace setters, the Operator is flagged as potentially vulnerable to insecure namespace-scoped resource references.

\mypar{Insecure Cluster-Scoped Resource Reference} 
This analysis aims to detect whether attacker-controlled input can influence cluster-scoped resources. Since cluster-scoped resources affect the entire Kubernetes cluster, any modification to them based on \namespace-scoped input represents a significant privilege escalation risk.
Thus, the analysis tracks data flow from the identified \namespace-scoped resource objects into any cluster-scoped resource objects.
If tainted input is used to construct cluster-scoped resource objects, the Operator is flagged as vulnerable to insecure cluster-scoped resource references.

\subsubsection{Summary}
To understand what an adversary can do to which kind of resource, the measurement further identifies insecurely referenced resource types and operations towards these resources. 

\mypar{Affected Resource Type Identification} This step discovers which resource can be referenced by an adversary. 
This identification is trivial for insecure cluster-scoped resource references, as their sink site in the vulnerability detection phase is set to cluster-scoped resource objects. Thus, the affected cluster-scoped resource types can be directly extracted from sink objects.

For insecure \namespace-scoped resource references, the affected resource type identification depends on the type of sink site in the previous step. If the previous data flow sinks at the three fields, \textit{WithNamespace()}, or \textit{SetNamespace()} methods, then the analyzer further tracks interprocedural data flow from the previous sinks to any \namespace-scoped resource objects to identify affected resource objects and types. If the previous data flow sinks at the constructor of a typed client, then the resource type is the one associated with that typed client. 

\mypar{Verb Identification}
To understand what an adversary can do to the insecurely referenced resources, the analyzer identifies the Kubernetes API Verbs (e.g., Get, Create, Update, Delete, etc.) related to insecurely referenced resources. If an insecurely referenced resource is found to be related to a Verb, like Create, then the adversary can exploit the vulnerable Operator to create the insecurely referenced resource in the Kubernetes cluster, which they should not.

Verb identification is achieved by identifying client method invocations that accept insecure references. 
For \textit{controller-runtime} library, it processes all resource types by a unified typeless client in the \textit{sigs.k8s.io/controller-runtime/pkg/client} package. For \textit{client-go} and \textit{client-gen} libraries, they process each type of resource with a specific typed client. Each Kubernetes API Verb corresponds to the client method with the same name. The analyzer thus performs interprocedural taint tracking from the reference site to these client methods to identify the related verbs.


\subsection{Measurement Result}
\label{subsec:measurement_result}
We conducted measurements on 2,268 Operators crawled from GitHub to assess the real-world impacts of insecure cross-\namespace references. The measurement suite was run on a Windows 11 machine with an Intel i7-10700K CPU (3.80GHz) and 32GB RAM. The suite cost 40026 seconds, with 17.6 seconds per Operator on average. In this part, we answered the following research questions:

\begin{itemize}[leftmargin=15pt]
\item \textbf{RQ1:} How many operators are potentially vulnerable to insecure cross-namespace reference?
\item \textbf{RQ2:} What resources can be cross-namespace referenced by attackers?
\item \textbf{RQ3:} What can attackers do towards cross-namespace referenced resources?
\item \textbf{RQ4:} How can insecure cross-namespace references impact the real world?
\end{itemize}

\subsubsection{RQ1: How Many Operators Are Potentially Vulnerable To Cross-Namespace Reference?}

\begin{figure}[ht]
    \centering
    \includegraphics[width=0.6\columnwidth]{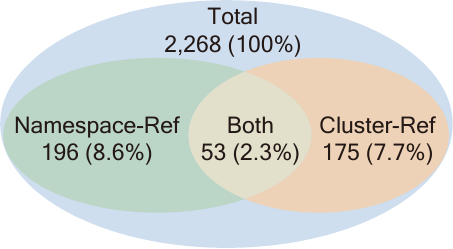}
    \caption{Percentage of Affected Operators}
    \label{fig:affect-ratio}
\end{figure}

To assess the prevalence of insecure cross-\namespace reference vulnerabilities, we analyzed a dataset of 2,268 real-world Kubernetes Operators collected from GitHub. Each Operator was examined to determine whether attacker-controlled \namespace-scoped resources can influence operations across namespace boundaries. The results illustrated that 318 Operators were potentially vulnerable, affecting 282 Operator providers. Specifically:
\begin{itemize}[leftmargin=15pt]
    \item 196 Operators (8.6\%) include insecure \namespace-scoped resource references that can specify or influence operations on other namespaces.
    \item 175 Operators (7.7\%) contain insecure cluster-scoped resource references, where attacker-controlled \namespace-scoped resources can affect cluster-scoped resources.
    \item 53 Operators (2.3\%) allow both types of references, posing risks of privilege escalation at both the namespace and cluster level.
\end{itemize}

These findings illustrate that a non-negligible portion (over 14\%) includes logic that may lead to privilege escalation, highlighting a widespread but largely overlooked security concern in the Kubernetes ecosystem.

\begin{table}[ht]
\centering
\caption{Operator Type Summary}
\label{tab:operator_type}
\begin{tabular}{@{}llll@{}}
\toprule
\textbf{Type}         & \textbf{Validated?} & \textbf{\# of Flagged} & \textbf{\# of Validated} \\ \midrule
Cluster   Management  & \checkmark          & 56            & 11                                     \\
Database              & \checkmark          & 52            & 9                                      \\
Networking            & \checkmark          & 34            & 6                                      \\
Cloud Management      & \checkmark          & 25            & 3                                      \\
CI/CD                 & \checkmark          & 23            & 3                                      \\
Metrics               & \checkmark          & 22            & 3                                      \\
Storage               & \checkmark          & 17            & 4                                      \\
Gadgets               & \checkmark          & 12            & 3                                      \\
Secret Management     & \checkmark          & 10            & 1                                      \\
Device                & \checkmark          & 9             & 2                                      \\
Workflow              & \checkmark          & 8             & 1                                      \\
Testing               & \checkmark          & 7             & 4                                      \\
Identity              & ×          & 7             & 0                                      \\
AI                    & \checkmark          & 7             & 2                                      \\
OS Management         & \checkmark          & 6             & 1                                      \\
Searching             & \checkmark          & 4             & 1                                      \\
Security              & ×          & 4             & 0                                      \\
Framework             & ×          & 3             & 0                                      \\
Config Management     & \checkmark          & 3             & 1                                      \\
CMS                   & ×          & 2             & 0                                      \\
Financial              & ×          & 2             & 0                                      \\
Registry              & ×          & 2             & 0                                      \\
Backup                & ×          & 1             & 0                                      \\
Collaboration         & ×          & 1             & 0                                      \\
Streaming             & ×          & 1             & 0                                      \\ \midrule
\textbf{Total} & \textbf{16}\checkmark & \textbf{318}  & \textbf{55}                            \\ \bottomrule
\end{tabular}
\end{table}

Table \ref{tab:operator_type} illustrates the type distribution of flagged Operators. These Operators cover a wide range of usage types in the real world, where Cluster Management, Database, and Networking are most prevalent.

We further assessed the accuracy of our measurement. Due to the lack of existing datasets or detection tools for this new class of vulnerabilities, it is infeasible to validate all Operators at scale manually. Thus, we primarily focused on false positives, evaluating if flagged Operators were truly vulnerable. We randomly reviewed 55 flagged Operators without any specific tailored criteria, covering major types of Operators as listed in Table \ref{tab:operator_type}. Although this sampling may not cover all potential Operator variations or directly conclude the perfect representative, it provides a practical basis to estimate the accuracy of our detection by covering the major types (or most prevalent types) of Operators, as we shown in Table~\ref{tab:operator_type}. 
Among the 55 cases, only 5 were false positives. This result indicates that the vast majority of cases flagged are indeed vulnerable, which supports the reliability of our measurement concerning the prevalence of vulnerability. While false negatives are difficult to quantify due to the lack of ground truth, our analysis focused on a range of commonly observed cross-namespace patterns, which may not capture all potential variations but still reflect realistic threats and led to confirmed vulnerabilities. Overall, the measurement provides strong evidence that Cross-\Namespace Reference Vulnerabilities are non-negligible in the real-world ecosystem.



\subsubsection{RQ2: What Resources Can Be Cross-Namespace Referenced By Attackers? }

\begin{table}[ht]
\centering
\caption{Major Insecurely Referenced Resource Type}
\label{tab:major-resources}
\begin{tabular}{@{}cll@{}}
\toprule
\multicolumn{1}{l}{\textbf{Scope}}                                                         & \textbf{Resource Type}                       & \textbf{Ref By \#Op.} \\ \midrule
\multirow{6}{*}{\begin{tabular}[c]{@{}c@{}}Namespace\\ \end{tabular}} & Secret                         & 102      \\
                                                                                 & ConfigMap                      & 29       \\
                                                                                 & Deployment                     & 29       \\
                                                                                 & Service                        & 22       \\
                                                                                 & StatefulSet                    & 12       \\ \midrule
\multirow{6}{*}{\begin{tabular}[c]{@{}c@{}}Cluster\\ \end{tabular}}   & Namespace             & 62       \\
                                                                                 & ClusterRoleBinding                      & 40       \\
                                                                                 & ClusterRole                    & 26       \\
                                                                                 & Node               & 25        \\
                                                                                 & PersistentVolume & 15        \\ \bottomrule
\end{tabular}
\end{table}

To understand the attack surface exposed by insecure cross-\namespace references, we investigate the types of resources that Operators allow attackers to reference across namespaces. For each resource type, we count the number of Operators that insecurely reference it and analyze which types are most frequently involved.

Among \namespace-scoped resources, the most commonly insecurely referenced types (listed in Table \ref{tab:major-resources}) are Secret (referenced by 102 Operators), ConfigMap (29 Operators), and Deployment (29 Operators). In Kubernetes, Secrets store highly sensitive data such as API keys, credentials, and TLS certificates. ConfigMaps often contain important application configurations that control applications' behavior, like API endpoints and performance arguments. Deployments define and manage the application workloads by controlling replica sets and pods.

For cluster-scoped resources, the most common insecurely referenced types are Namespace (62 Operators), ClusterRoleBinding (40 Operators), and ClusterRole (26 Operators). In Kubernetes, Namespace is the resource that defines a namespace in a Kubernetes cluster. ClusterRoles and ClusterRoleBindings define and grant cluster-level permissions that apply to the whole cluster.

\begin{figure}[ht]
    \centering
    \includegraphics[width=\columnwidth]{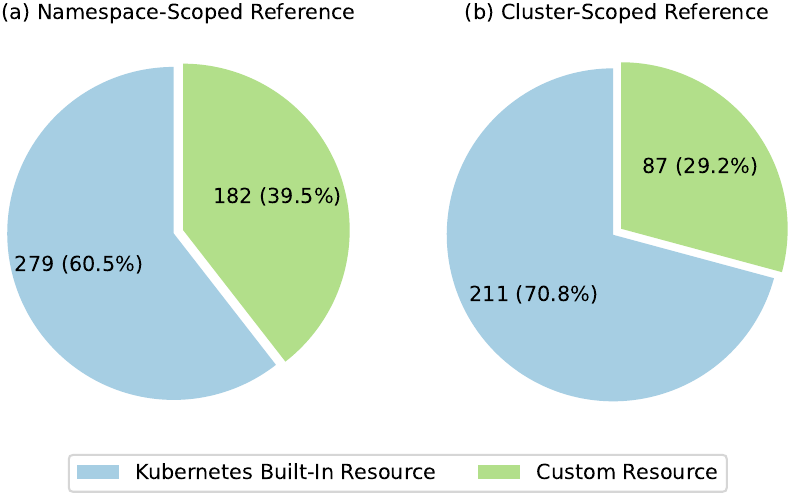}
    \caption{Reference of Built-In and Custom Resources}
    \label{fig:resource-type}
\end{figure}

We also distinguish between insecure references to built-in Kubernetes resources and custom resources. To this end, we aggregated the type-\#Operator result above based on built-in resource type or custom resource type. For insecure \namespace-scoped references, 279 cases involved built-in resources and 182 involved custom resources. For insecure cluster-scoped references, 211 targeted built-in resources and 87 involved custom resources. These results indicate that insecure references can affect both built-in resources and custom resources. And the insecure references are more commonly associated with Kubernetes built-in resources.

\subsubsection{RQ3: What Can Attackers Do Towards Cross-Namespace Referenced Resources?}

\begin{figure}[ht]
    \centering
    \includegraphics[width=\columnwidth]{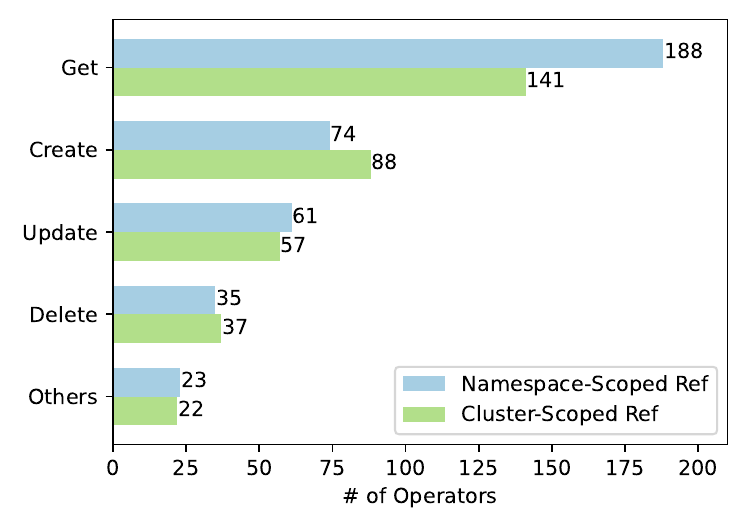}
    \caption{Verbs Used By \# Operators Towards Insecurely Referenced Resources}
    \label{fig:verbs}
\end{figure}

To understand the potential impact of insecure cross-\namespace references, we analyze the operations (Kubernetes API Verbs) that Operators perform on the referenced resources. For each identified insecurely referenced resource, we extract its related verbs. We then count how many Operators apply each verb to each insecurely referenced resource type. The result is illustrated in \autoref{fig:verbs}.

For insecurely referenced \namespace-scoped resources, the top three most common verbs are \textit{Get} (used by 188 Operators), \textit{Create} (74 Operators), and \textit{Update} (61 Operators), with \textit{Get} being the most prevalent. In Kubernetes, \textit{Get} retrieves the current state of a resource, \textit{Create} instantiates a new resource, and \textit{Update} modifies an existing resource. The predominance of the \textit{Get} operation indicates that a large number of vulnerable Operators retrieve data from resources in other namespaces based on attacker-controlled inputs, exposing unauthorized data to attackers.

For insecurely referenced cluster-scoped resources, the top three verbs are \textit{Get} (used by 141 Operators), \textit{Create} (88 Operators), and \textit{Update} (57 Operators), with \textit{Get} accounting for the largest proportion. This suggests that in many cases, Operators may use attacker-influenced data to get cluster-wide resources, which may expose sensitive cluster-level information to attackers.

\begin{figure}[ht]
    \centering
    \includegraphics[width=\columnwidth]{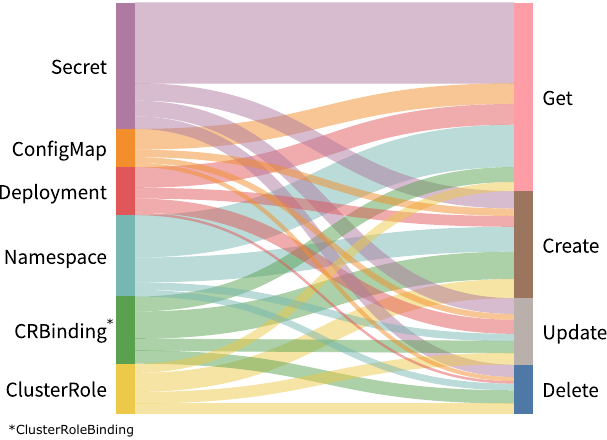}
    \caption{Verbs of Major Insecurely Referenced Resources}
    \label{fig:all-corr}
\end{figure}



\begin{table}[ht]
\centering
\caption{Major Insecurely Used Verb-Resource Pairs}
\label{tab:resource-verb}
\begin{tabular}{cll}
\toprule
\multicolumn{1}{l}{\textbf{Scope}}                                                         & \textbf{Verb - Ref.Resource Type}               & \textbf{Used By \#Op.} \\ \midrule
\multirow{6}{*}{\begin{tabular}[c]{@{}c@{}}Namespace\end{tabular}} & Get - Secret                & 97          \\
                                                                                 & Get - ConfigMap             & 25          \\
                                                                                 & Get - Deployment            & 25          \\
                                                                                 & Create - Secret             & 21          \\
                                                                                 & Update - Secret             & 19          \\ \midrule
\multirow{6}{*}{\begin{tabular}[c]{@{}c@{}}Cluster\end{tabular}}   & Get - Namespace & 51          \\
                                                                                 & Create - ClusterRoleBinding          & 33          \\
                                                                                 & Create - Namespace        & 29          \\
                                                                                 & Create - ClusterRole & 23          \\
                                                                                 & Get - Node        & 20           \\ \bottomrule
\end{tabular}
\end{table}

To understand the practical implications, we investigate the verb-resource pairs (i.e., combinations of verbs and insecurely referenced resource types) to determine which verbs Operators typically perform on specific insecurely referenced resource types. The result is illustrated in Table \ref{tab:resource-verb} and \autoref{fig:all-corr}.

For insecurely referenced \namespace-scoped resource types, the most prevalent pairs are \textit{Get-Secret} (used by 97 Operators), \textit{Get-ConfigMap} (25 Operators), and \textit{Get-Deployment} (25 Operators). These indicate that a substantial number of Operators can be exploited to read data from Secrets, configuration files, or application deployments in attackers' unauthorized namespaces, causing information exposure.

For insecurely referenced cluster-scoped resource types, the most prevalent pairs are \textit{Get-Namespace} (used by 51 Operators), \textit{Create-ClusterRoleBinding} (33 Operators), and \textit{Create-Namespace} (29 Operators). These patterns suggest that Operators may be exploited to reveal other namespaces in the cluster, assign cluster-wide permissions, or provision new namespaces, leading to privilege elevation.

Together, these findings highlight that insecure cross-\namespace references are not only present but often tied to high-impact operations on sensitive or privileged Kubernetes resources.

\subsection{RQ4: Case Study}
To validate the practical significance of the identified vulnerabilities, we conducted in-depth inspections of vulnerable Operators. Combined with static analysis and manual exploits, the detailed impacts of their vulnerabilities are identified. We responsibly disclosed our findings to their vendors. By the time of submission, 8 vulnerabilities have been confirmed by the affected vendors as listed in Table \ref{tab:confirmed-vulns}, where all vulnerable resources are custom resources, as Operators are typically triggered by their custom resources and then reference other resources. 7 CVEs have either been assigned or are currently under processing, reflecting community acknowledgment and real-world relevance of these issues. 
Notably, even Operators maintained by Red Hat, the inventor of Operators, were confirmed to be vulnerable to insecure cross-\namespace references, highlighting that the problem is systemic and not limited to less mature projects.

We introduced the case of Grafana/tempo-operator to illustrate how these insecure references impact the real world.
\begin{table*}[ht]
\centering
\caption{Vendor-Confirmed Vulnerabilities}
\label{tab:confirmed-vulns}
\begin{threeparttable}
\begin{tabular}{@{}llllll@{}}
\toprule
\textbf{Operator}                & \textbf{Vendor}             & \textbf{Status}               & \textbf{Vulnerable Resource}   & \textbf{Referenced Resource} & \textbf{Verb}   \\ \midrule
Operator X \tnote{1} & ******             & Confirmed & ******        & ******              & ******    \\
gateway-operator       & Kong               & CVE Assigning        & AIGateway       & Secret              & Get \\
baremetal-operator         & Metal3-io            & CVE-2025-29781       & BMCEventSubscription        & Secret              & Get    \\
observability-operator & Red Hat            & CVE-2025-2843        & MonitorStack    & ClusterRoleBinding  & Create/Update/Delete \\ 
gateway-operator       & Kong               & CVE Assigning        & ControlPlane    & ClusterRoleBinding  & Create    \\
tempo-operator         & Grafana \& Red Hat & CVE-2025-2842        & TempoStack      & ClusterRoleBinding  & Create/Delete \\
tempo-operator         & Grafana \& Red Hat & CVE-2025-2786        & TempoMonolithic & ClusterRoleBinding  & Create/Delete \\ 

ais-k8s             & NVIDIA             & CVE-2025-23260
            & AIStore         & ClusterRoleBinding  & Create/Update/Delete \\ \bottomrule
\end{tabular}
\begin{tablenotes}
\item[1]  Hidden for Ethical Consideration. 
\end{tablenotes}
\end{threeparttable}
\end{table*}

Grafana \cite{cncf}, a member of CNCF, is a widely adopted open-source analytics and visualization platform, renowned for transforming complex data into interactive dashboards. Among their products, Grafana Tempo \cite{tempo-intro} stands as a distributed tracing backend for the Grafana Observability Stack, gaining over 4,000 stars on GitHub.
The tempo-operator \cite{tempo}, developed by both Grafana and Red Hat, is the official solution for deploying and managing Grafana Tempo on Kubernetes clusters. It defines a \namespace-scoped Custom Resource, \textit{TempoStack}, for users to deploy Tempo.

Specifically, when enabling \textit{JaegerQuery} and \textit{MonitorTab} functions in a \textit{TempoStack} resource, tempo-operator would create a ClusterRoleBinding, granting cluster-level permissions to a ServiceAccount in the namespace specified by the attacker-controlled \textit{TempoStack.Namespace} field. 
An attacker can thus deploy a \textit{TempoStack} in his namespace, where the \textit{JaegerQuery} and \textit{MonitorTab} functions were enabled and \textit{TempoMonolithic.Namespace} field was set to the name of his authorized namespace. The Operator will then grant cluster-level permissions to a ServiceAccount in his namespace. The attacker can then impersonate the ServiceAccount to gain access to resources in the whole cluster. 

In terms of impacts, the operator would grant the \textit{cluster-monitoring-view} ClusterRole, an OpenShift-specific role enabling access to Prometheus and Thanos APIs, thus allowing the attacker to observe metrics and monitor cluster-wide resource states. We validated the bug and reported it to the vendors. They responded that \textit{``We confirmed the vulnerability and will start the process to assign a CVE shortly. ''}. CVE-2025-2842 was then assigned to us.


\section{Mitigation \& Discussion}

During our inspection of vulnerable Operators, we realized that Operators aim to simplify user operations as much as possible, thus may embed cross-\namespace reference functionality to spare users the repetitive, manual task of duplicating resources like Secrets across namespaces—a practice necessitated by Kubernetes' namespace isolation. However, this convenience can inadvertently introduce vulnerabilities if not properly implemented, as attackers may exploit such helpful behavior to perform unauthorized cross-\namespace actions and privilege escalations.
Thus, we suggest the following mitigations to eliminate the cross-\namespace reference vulnerability.


\mypar{Carefully Using Multi-Tenant Kubernetes} 
Practitioners are discouraged from sharing Kubernetes across untrusted users, ensuring all tenants in multi-tenant Kubernetes do no evil. 
Developers should eliminate vulnerabilities of applications running on multi-tenant clusters, which may be exploited to gain initial Kubernetes access and affect other tenants.

\mypar{Scope Alignment}
Developers are advised to ensure that the declared scope of resources accurately reflects the scope of their operational effect. If a resource is defined as \namespace-scoped but its process logic performs actions across multiple namespaces at the cluster level, it creates a dangerous mismatch between the resource’s access control boundary and its actual impact. In such cases, the resource should be explicitly declared as cluster-scoped, ensuring that only privileged users can create or manipulate it. 



\mypar{Mitigate with Kubernetes Admission Control}
In cases where cross-\namespace is necessary or modifying Operators is painful, we designed a mitigation approach based on \textit{Validating Admission Webhook} \cite{admissioncontrol} to reject harmful requests before proceeding to the original Operators. The mitigation works independently, demands no modification to existing Operators, and can be easily adapted to any project.


Essentially, Validating Admission Webhook is a mechanism to extend the native Kubernetes access control. When a user manipulates a resource, besides RBAC, Kubernetes will also invoke Webhooks to decide whether to allow or reject the request. In addition, Kubernetes provides the \textit{SubjectAccessReview} \cite{subjectaccessreview} API, which can be used to check whether a user has specific RBAC permissions on a given resource. We leverage these mechanisms to further check whether the user has permissions for referenced resources and inform Kubernetes to reject requests if the user is unauthorized. This fills the gap of the native Kubernetes RBAC, which is not aware of whether a resource has cross-\namespace references.

\begin{figure}
    \centering
    \includegraphics[width=0.81\linewidth]{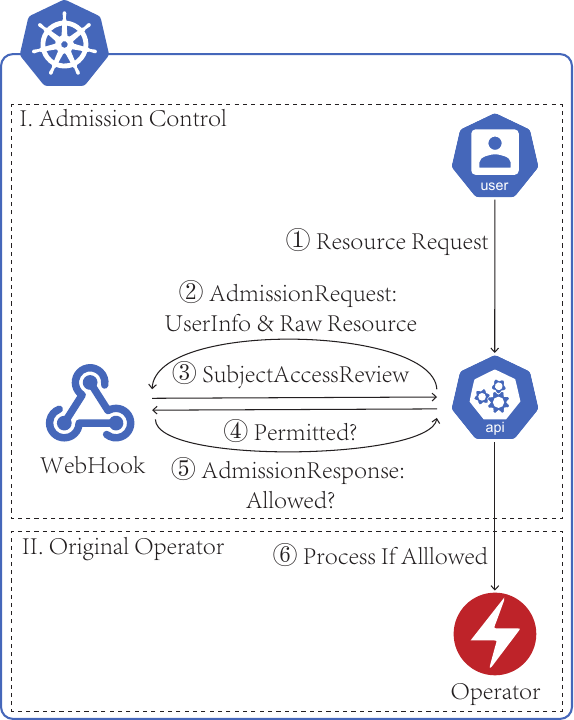}
    \caption{Mitigate with Validating Admission Webhook}
    \label{fig:webhook}
\end{figure}

\autoref{fig:webhook} gives the workflow of our proposed mitigation. The basic flow of a Validating Admission Webhook is that the Kubernetes API Server intercepts a user request, sends an \textit{AdmissionReview} object containing the raw resource and user identity to the Webhook, and waits for an \textit{AdmissionResponse} object from the Webhook indicating whether the request should be accepted or rejected.
Upon receiving an \textit{AdmissionReview} from the Kubernetes API Server, the Webhook reads the object and evaluates whether the user has cross-namespace permissions by issuing a \textit{SubjectAccessReview} query to the API Server. If the result indicates that the user lacks permissions, the Webhook responds with an \textit{AdmissionResponse} with the \textit{allowed} field set to \textit{false}, causing the API Server to reject the request before it reaches the Operator.

\begin{figure}
    \centering
    \includegraphics[width=0.90\linewidth]{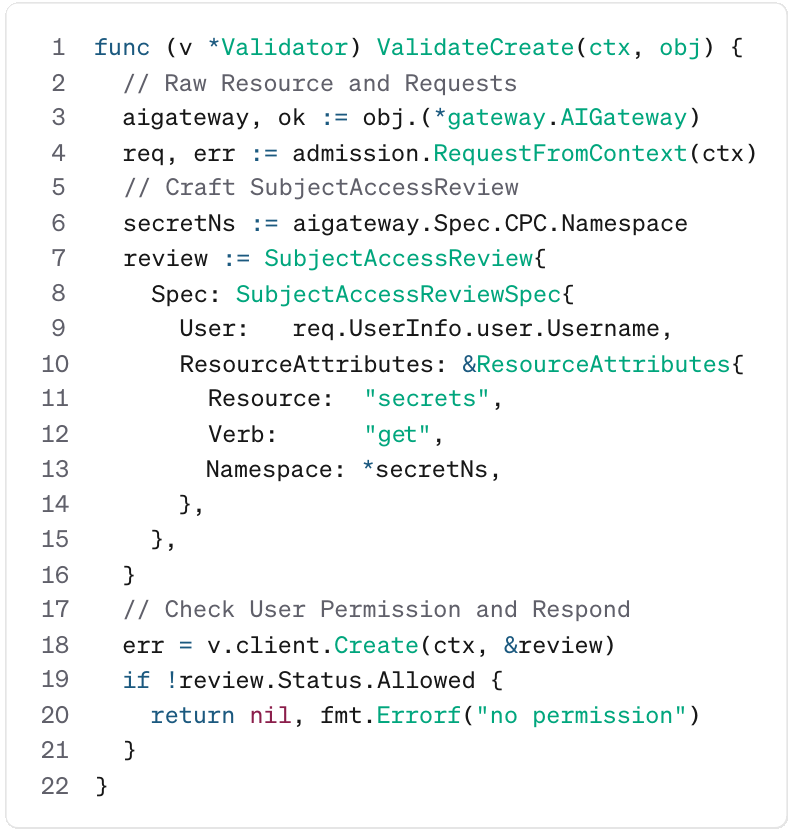}
    \caption{Simplified Admission WebHook Example For Mitigating gateway-operator Vulnerability}
    \label{fig:webhook-example}
\end{figure}

We implemented a Validating Admission Webhook with Kubebuilder \cite{kubebuilder}. \autoref{fig:webhook-example} illustrates a simplified version of our Webhook to mitigate the cross-\namespace secret reference vulnerability of the Kong gateway-operator. The Line 3-4 extracts the raw resource and raw request. The Line 6 gets the namespace of the referenced secret from the raw resource. The Line 7-16 crafts a \textit{SubjectAccessReview} query, which checks whether the requesting user has permission to Get Secrets in the referenced namespace. The Webhook issues the query at Line 18, checks the result at Line 19, and rejects the unauthorized user at Line 20. With the scaffolding generated by Kubebuilder, tens of lines of code are required to implement a basic mitigation. And developers or cluster administrators can further implement other finer-grained validations in complex scenarios if needed.
\begin{table}[ht]
\centering
\caption{Validating Admission Webhook Mitigation Evaluation}
\label{tab:mitigation}
\begin{threeparttable}
\begin{tabular}{@{}lllr@{}}
\toprule
\textbf{Operator}       & \textbf{Mitigated} & \textbf{Before (ms)} & \textbf{After (ms)} \\ \midrule
Operator X & \color{ForestGreen}\faCheckCircle     & 95.7                 & 99.3 (+3.6)         \\
gateway-operator        & \color{ForestGreen}\faCheckCircle     & 94.5                 & 97.0 (+2.5)           \\
baremetal-operator            & \color{ForestGreen}\faCheckCircle     & 91.6                 & 94.5 (+2.9)         \\
observability-operator  & \color{ForestGreen}\faCheckCircle     & 96.5                 & 101.2 (+4.7)        \\
tempo-operator          & \color{ForestGreen}\faCheckCircle     & 106.3                & 113.7 (+7.4)        \\
ais-k8s           & \color{ForestGreen}\faCheckCircle     & 101.7                & 105.2 (+3.5)        \\ \midrule
\multicolumn{2}{l}{\textbf{Average}}         & 97.7                 & 101.8 (+4.1)        \\ \bottomrule
\end{tabular}
\end{threeparttable}
\end{table}

We evaluate its effectiveness and overhead on the vulnerable Operators listed in Table~\ref{tab:mitigation}. The evaluation is conducted on a Kubernetes cluster with 1 control plane node and 3 worker nodes. Each node is equipped with 8 cores of Intel E5-2680 v4 CPU (2.40GHz) and 32GB RAM. To evaluate the effectiveness, we created two ServiceAccounts in the cluster: one with cluster-level permissions and another limited to a single namespace. Before deploying the mitigation, both ServiceAccounts were able to invoke all listed vulnerable Operators to perform cross-\namespace references. After deploying the mitigation, only the ServiceAccount with legitimate cross-\namespace permissions could invoke the affected Operators, while the ServiceAccount with single-\namespace permissions was denied in all cases. This demonstrates that the mitigation successfully enforces namespace boundaries and prevents unauthorized privilege escalation.

To evaluate the overhead, since the Webhook only works at admission time and operates independently before the Operator logic, we measured the time between the point at which a new resource request was initiated and the point the Kubernetes API Server accepted the resource. For each test case, we repeated it 10 times and calculated the average overhead. Experimental results show an average increase of  4.1~ms after applying the mitigation. Given that the added overhead is a one-time overhead for a single resource request and Operator logic typically takes seconds to minutes, this added delay is negligible in practice.

To facilitate the mitigation, we open-sourced our Webhook for reference and a lightweight automatic Webhook generator, enabling quick adoption for practitioners.




\mypar{Difference with Related Kubernetes Attacks}
The cross-\namespace attack is fundamentally different from existing attacks \cite{takeover, k8sblackhat, podescape} regarding threat models, root causes, and defenses.

For threat models, as illustrated in Section \ref{sec:threatmodel}, the precondition of launching existing attacks is to assume attackers have compromised containers (e.g., get a shell) before conducting overprivilege attacks, leaving critical gaps in how to compromise containers. In contrast, we presented a practical threat model without such strong assumptions, detailing two practical strategies to exploit Operators from scratch.

For root causes, existing attacks stem from granting applications unused permissions. Once attackers compromise applications (e.g., gain shells), they can conduct existing attacks by using over-granted privileges in follow-up actions. Thus, they \cite{takeover, k8sblackhat, podescape, redhatreport, operatorsynk, operatorkubeops} suggest the Principle of Least Privilege (PoLP). Since only removing unused permissions, PoLP will not affect their functionality and is an ideal defense.

In contrast, our attack arises even when Operators follow PoLP. Supposing an Operator reading user input to get Secrets in a specified namespace (rather than assuming gain shells), it typically implies the Operator needs secret information for its operation, thus requiring certain privileges to function correctly. Removing such privileges would break the Operator’s functionality, even for legitimate intra-\namespace references.
In other words, our vulnerability stems from insecure use of legitimate privilege, which existing defenses cannot address. An ideal defense should preserve necessary privileges while mitigating vulnerabilities; therefore, we propose new mitigations.

\mypar{More Cross-Tenant Attacks}
Similar cross-tenant attacks can also happen in various cloud scenarios beyond Kubernetes.
Security researchers have discovered multiple services provided by Microsoft Azure \cite{ChaosDB, extrareplica, synapse, ACSESSED, entraid}, AWS \cite{awsecr, appsync}, Google Cloud Platform \cite{gcpbigquery, gcplookstudio}, and Oracle \cite{attachme}
were vulnerable to cross-tenant attacks, allowing attackers to manipulate unauthorized resources of other tenants. Our work further extends the landscape to Kubernetes and emphasizes cross-tenant risks in modern cloud architectures.

\section{Related Work}


\mypar{Kubernetes Operators and Controllers} Existing research on Kubernetes operators and controllers focuses on functional bugs instead of attacks.
Acto \cite{acto} proposes an automatic end-to-end testing technique for validating the operational correctness of Kubernetes Operators. Acto continuously generates desired state declarations and verifies whether the Operator correctly reconciles the system to those states. 
Sieve \cite{sieve} presents an automatic reliability testing framework for cluster-management controllers. By injecting faults, Sieve uncovers deep semantic bugs by observing how controllers behave under fault conditions they are expected to tolerate. 
Anvil \cite{anvil} presents the formal verification framework for Kubernetes controllers via TLA-style temporal reasoning, validating whether controllers eventually bring the cluster to the desired state and maintain it. 
Kivi \cite{kivi} verifies Kubernetes controllers and their configurations by modeling controller behaviors and checking for violations of user-defined intent properties using model checking. It detects issues like imbalance and lifecycle bugs, focusing on functional correctness.
\citet{operatorbug} systematically summarize historical functional bugs of Operator.  
Red Hat \cite{operatorsecpractice}, Synk \cite{operatorsynk}, and KubeOps \cite{operatorkubeops} suggest several good practices for Kubernetes Operators. However, their core guidance, limiting RBAC scope, is a general Kubernetes practice, not specific to Operator, and does not mitigate our proposed attack.

To the best of our knowledge, our work is the first comprehensive study on Kubernetes Operator attacks.

\mypar{Kubernetes Security}
In terms of attack and exploitation techniques, MITRE \cite{mitrematrix} and Microsoft \cite{msmatrix} summarize tactics to compromise containers and container orchestration systems like Kubernetes. \citet{kubedevopsprivesc} investigate privilege escalation scenarios for DevOps pipelines on Kubernetes. 
\citet{ziyi} present cross-container attacks on Kubernetes with eBPF.  Page Spray attack~\cite{pagespray} can lead to container escaping, which can be mitigated by memory safety hardening and repair~\cite{camp,patchagent}. \citet{paced} design Cross-Linux-Namespace defenses to detect container escape attacks. \citet{honeypot} set up honeypots on Kubernetes and analyzes the attacks towards containers and container orchestration systems. \citet{coresidency} present a co-residency attack towards container orchestration systems. \citet{qingyang} comprehensively analyze 30 vulnerabilities in Kubernetes stacks. \citet{k8sblackhat, podkubecon} identify the threat of trampoline pods which can be leveraged to gain escalated privileges.
However, these works have not revealed or addressed the security issues brought about by Kubernetes operators.

Kubernetes offers extensive configuration options for managing applications, including access controls and specifying security contexts. Any misconfigurations can lead to severe security vulnerabilities.
Thus, another theme of Kubernetes security research is eliminating misconfiguration. Shamim et al. \cite{xipractice, configurationsurvey, mitigatemisconfig} systematically reveal the risks of misconfiguration regarding best practice. 
\citet{kubemisconfigmeasure} design static analysis tools and conduct a large-scale empirical study on Kubernetes manifests, revealing the landscape of misconfiguration. 
\citet{kgsecconfig} leverage knowledge graphs to detect and mitigate Kubernetes misconfiguration.
Recent work \citet{takeover} identify the security risk of excessive Kubernetes RBAC permissions, which may lead to whole cluster takeover. EPScan \cite{epscan} follows up on the research and designs systems to automatically minimize RBAC permissions.
The industry also presents numerous tools for Kubernetes security, including Trivy \cite{trivy}, Kubescape \cite{kubescape}, KubeSec \cite{kubesec}, KubeArmor \cite{kubearmor}, Open Policy Agent \cite{opa}, and Kyverno \cite{kyverno}, providing functions like misconfiguration detection and runtime policy enforcement.

While the existing works try to address misconfiguration and achieve the Principle of Least Privilege (PoLP) for applications, the vulnerability we present is not simply misconfigurations or violations of PoLP. They arise from inherent flaws in how Operators process user-controlled resources. These vulnerabilities exist even when the permissions of Operators are minimal, highlighting a deeper design-level security gap in the Operator model itself.

\section{Conclusion}

In this paper, we presented the first in-depth research on Kubernetes Operator attacks, unveiling a long-neglected Cross-Namespace Reference Vulnerability with two strategies, demonstrating how an attacker can bypass namespace isolation. We designed and implemented a static analysis suite to conduct large-scale measurements, illustrating that over 14\% of Operators in the wild are potentially vulnerable. Our findings have been reported to the relevant developers, resulting in 8 confirmations and 7 CVEs by the time of submission, highlighting the critical need for enhanced security practices in Kubernetes Operators. We proposed concrete mitigation solutions and open-sourced our code to benefit the ecosystem.

\section*{Ethics Considerations}
We conducted our research with strict adherence to ethical guidelines. To collect Operator projects, we followed GitHub’s rate limits and usage policies. We validated vulnerabilities on our own Kubernetes clusters, ensuring that no third-party users or environments were affected. We responsibly disclosed vulnerabilities to all affected vendors. Thus, each affected project has at least 90 days to fix before publication. All vulnerability disclosures complied with the security policies of the respective vendors. All vulnerabilities in Table \ref{tab:confirmed-vulns} were validated and then reported to the vendors according to their vulnerability policies. To facilitate mitigation, we suggested available approaches, released mitigation samples and a lightweight mitigation generator, enabling quick adoption for practitioners.  We will ensure all vulnerabilities discussed in the case studies are fixed or authorized to mention by the time of the final publication. We are actively coordinating with vendors and will publish full datasets at the proper time (expected conference date) to ensure developers have enough time to fix and minimize impacts.

\section*{Acknowledgment}
We thank our reviewers and shepherd for their valuable feedback and comments. For authors, Ziyi Guo is supported by Google PhD Fellowship. Zhenyuan Li is supported by the National Natural Science Foundation of China under 62402419  and by the CCF-Tencent Rhino-Bird Open Research Fund. Any opinions, findings, and conclusions or recommendations expressed in this work are those of the author(s) and do not necessarily reflect the views of the institutions above.

\bibliographystyle{IEEEtranN}
\bibliography{references}
\end{document}